\documentclass[prl,aps,twocolumn,10pt,superscriptaddress,notitlepage,longbibliography]{revtex4-2}

\usepackage[dvipsnames]{xcolor}
\usepackage[english]{babel} 
\usepackage{graphicx}
\usepackage{braket}
\usepackage{epstopdf}
\usepackage{mathtools}
\usepackage{amsmath}
\usepackage{amssymb}
\usepackage{bbm,bm}
\usepackage[caption=false]{subfig}
\usepackage{pythonhighlight}
\usepackage{hyperref}
\usepackage{float}
\usepackage{bbold}
\usepackage[T1]{fontenc}
\usepackage[capitalize]{cleveref}

\usepackage{microtype}

\usepackage{dsfont}
\usepackage{bm}

\usepackage{tikz}

\usetikzlibrary{
    arrows.meta,
    positioning,
    calc,
    decorations.pathmorphing,
    shapes.geometric,
    fit
}

\graphicspath{{./figures/}{./Figures/}{./}}

\usepackage[markup=underlined]{changes}
\makeatletter
\@namedef{Changes@AuthorColor}{magenta}
\colorlet{Changes@Color}{magenta}
\makeatother
 \hypersetup{
     colorlinks=true,
     linkcolor=blue,
     filecolor=blue,
     citecolor = magenta,      
     urlcolor=red,
     }

\usepackage{xargs}
\newcommandx{\greencom}[2][1=]
{\todo[inline, color=green!40,#1]{#2}}
\newcommandx{\bluecom}[2][1=]
{\todo[inline, color=blue!40,#1]{#2}}
\newcommandx{\bluemargin}[2][1=]
{\todo[color=blue!40,#1]{#2}}

\usepackage{letltxmacro}
\LetLtxMacro{\ORIGselectlanguage}{\selectlanguage}
\makeatletter
\DeclareRobustCommand{\selectlanguage}[1]{%
  \@ifundefined{alias@\string#1}
    {\ORIGselectlanguage{#1}}
    {\begingroup\edef\x{\endgroup
       \noexpand\ORIGselectlanguage{\@nameuse{alias@#1}}}\x}%
}
\newcommand{\definelanguagealias}[2]{%
  \@namedef{alias@#1}{#2}%
}

\makeatother

\definelanguagealias{en}{english}
\definelanguagealias{EN}{english}
\definelanguagealias{eng}{english}
\definelanguagealias{de}{ngerman}

\newcommand{\gp}{\gamma_\phi}
\newcommand{\Ex}{\mathbb{E}}
\newcommand{\Gtt}{G^{(2)}_{TT}}
\newcommand{\Gba}{G^{(2)}_{ba}}
\newcommand{\Gttfac}{G^{(2),\mathrm{fac}}_{TT}}

\begin{document}

\title{Pure dephasing destroys two-photon amplitude, not nonlinear phase}

\author{M. Kozma}
\author{S. Hughes}
\affiliation{Department of Physics, Engineering Physics and Astronomy,
Queen's University, Kingston, Ontario K7L 3N6, Canada}

\date{\today}

\begin{abstract}
A two-level emitter chirally coupled to a waveguide imprints a $\pi$ conditional
phase on a two-photon pulse, a key component assumed in quantum photonic neural
network architectures, but whether it survives the pure dephasing of a
solid-state emitter is unknown, since no two-photon scattering treatment
includes it. We derive the dephased two-photon amplitude in closed form and
validate it against an independent Fock-state master equation. The conditional
phase is pinned at exactly $\pi$ at every dephasing rate examined: dephasing destroys
the amplitude carrying the nonlinear phase rather than rotating it. The one- and
two-photon channels are not attenuated equally, so dephasing does not 
simply
scale the node down but reweights it between photon-number
sectors---an error no
phase setting can train away. 
The standard factorization of two-photon
correlators, exact for arbitrary loss,
fails at first order in $\gp$, discarding
$16$--$81\%$ of pairs over $\gp/\gamma=0.1$--$1$.

\end{abstract}

\maketitle

A two-level emitter chirally coupled to a waveguide is the simplest optical nonlinearity that acts
at the level of single
photons~\cite{ShenShen2015,Chang_2012,Lodahl2015InterfacingNanostructures,LeJeannic2022,PhysRevX.10.031011},
and in the ideal limit it realizes a full $\pi$ conditional phase between one- and two-photon
inputs---the ideal Kerr response, and twice the $\pi/2$ ceiling of a bidirectional
geometry~\cite{Lodahl2017,Petersen2014,Sllner2015,Coles2016,MartinCano2019,JalaliMehrabad:20}.

This is no longer an idealization in practical photonic systems: Staunstrup
\emph{et al.}~\cite{Staunstrup2024} have measured the emitter-induced phase
interferometrically in a bidirectional waveguide, and a rapidly growing body of
architecture work is built on exactly this response: quantum photonic neural
networks~\cite{Ewaniuk2023,Ewaniuk2025,QPNN2026}, chiral 
 quantum networks~\cite{PhysRevLett.117.240501} and cluster-state
generators~\cite{Istrati2020}, the most
recent modeling the emitter with its full chiral scattering
matrix~\cite{QPNN2026}, but without pure dephasing. Continuous-wave chiral nonlinear optics and
all-optical control draw on the same
interaction~\cite{Dufresne2026a,Dufresne2026b}.
In a bidirectionally coupled waveguide the phase excursion is bounded by $\pi/2$ and
maximal off resonance; the measurement gives $\varphi_{\max}=-0.19\pi\pm0.03$,
and Ref.~\cite{Staunstrup2024} identifies chiral coupling as the route to the
full, on-resonance $\pi$, which is the regime treated here. Quantum emitters
also serve as reconfigurable phase elements in otherwise linear
circuits~\cite{McCaw2024}.

The difficulty is that \emph{all solid-state emitters dephase}.
Inverting the measured
phase excursion on the two dipoles of Ref.~\cite{Staunstrup2024} gives
total linewidths $1.4$ and $3.2$ times the radiative rate, i.e., pure dephasing
rates of order $\gp/\gamma \sim 0.3$ and ${\sim}2$ for a typical coupling
efficiency~\cite{Lund-Hansen2008,10.1063/1.5117888}. These are not perturbative
corrections. More generally, $\gp/\gamma\simeq0.1$--$1$ is the range that matters
in practice: lower rates have been reached in diode-embedded structures where
background charge noise is suppressed electrically~\cite{Kuhlmann2013,pedersen2020_noise}, but such devices are
demanding to integrate at the scale a network requires, and
$\gp/\gamma\gtrsim0.1$ is typical of state-of-the-art emitters in waveguides.

Yet the theory available to
describe them is {\it incomplete} in a specific way. Frequency-domain two-photon 
scattering theory, the common language of this
field~\cite{ShenShen2015,Nysteen2017,Ewaniuk2025,PhysRevLett.126.023603,Trschmann2019}, has no dephasing implementation. 
Matrix-product-state and related non-Markovian
methods~\cite{PhysRevLett.116.093601,PhysRevResearch.3.023030,Yanagimoto2021,PhysRevResearch.3.023168,Strathearn2018EfficientOperators,PhysRevLett.128.167403,Matias_2026,githubQwaveMPS}
do include it, and Bundgaard-Nielsen \textit{et al.}~\cite{Matias_2026} recently
gave the single-photon results in full. Their analytic treatment stops there,
however, and they state explicitly that it ``fails here'' for two photons, which
is the gate-relevant case, since a nonlinear phase gate is defined by the
\emph{difference} between one- and two-photon response rather than by either alone.

In this Letter, 
we derive the two-photon scattered amplitude of a chirally coupled emitter driven by a
Fock pulse, with pure dephasing included exactly, in closed form, validated against a
Fock-state master equation solved with no factorization. Three key
results follow: 
(i) the conditional phase is exactly $\pi$ at \emph{every} dephasing rate
examined, $\gp/\gamma=0$ to $2$, including above the radiative rate where the
single-photon phase has flipped to $0$; dephasing costs amplitude, not phase;
(ii) the one- and two-photon channels do not attenuate together, so dephasing
does not simply dim the node's output---it changes the balance between its
one- and two-photon responses, which alters what the node does to any input containing both;
(iii) 
the factorization of two-photon correlators used throughout the scattering
literature is exact for arbitrary loss but fails at first order in $\gp$, and we
give the exact replacement.

\begin{figure}[t]
\centering
\includegraphics[width=\columnwidth]{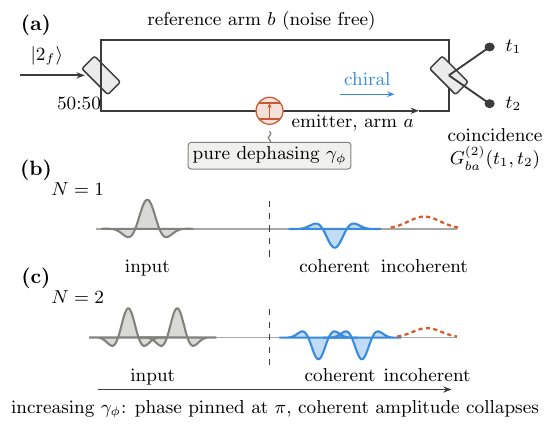}
\caption{(a) A two-level emitter chirally coupled to a waveguide, subject to pure
dephasing at rate $\gp$. A Fock pulse is split on a 50:50 beamsplitter into a
noise-free reference arm $b$ and a signal arm $a$ containing the emitter; the
arms are recombined and coincidences at times $t_1$ and $t_2$ give the
interferometric correlator $\Gba$ that measures the phase. (b),(c) One- and
two-photon Fock pulses scatter into a coherent output carrying the $\pi$ shift
(drawn inverted) whose amplitude shrinks with $\gp$, plus an incoherent
remainder that carries the missing weight. Photon number is conserved at every
$\gp$; coherent amplitude is not.}
\label{fig:schematic}
\end{figure}

\textit{Model.}---We take a two-level system, $\sigma^-=\ket{g}\bra{e}$, coupled
to a chiral one-dimensional continuum within the rotating-wave and first Markov
approximations. The emitter decays into the guided mode at rate
$\gamma\equiv\gamma_R$ and into all other channels at $\gamma_0$, so
$\Gamma=\gamma+\gamma_0$ and $\Gamma_2=\Gamma+\gp$, and we define
$\zeta=i\Delta+\Gamma_2/2$ with 
$\Delta=\omega_0-\omega_p$ 
the emitter--pulse detuning. 
Time-domain amplitudes are written
in the pulse frame, so the detuning enters through $\zeta$, while
frequency-domain quantities such as $t(\omega)$ are referred to $\omega_0$.

The input--output relation is
{$a_{\rm out}(t)=a_{\rm in}(t)-i\sqrt\gamma\,\sigma^-(t)$}. All results below are
quoted for the ideal chiral case $\gamma_0=0$, so $\Gamma=\gamma$, though
Eq.~\eqref{eq:fp} holds for arbitrary $\gamma_0$~\footnote{Because $\gamma_0$
and $\gp$ enter the coherent amplitudes only through $\Gamma_2$, the coherent
output of a dephased emitter is identical to that of an emitter with off-chip
decay $\gamma_0+\gp$ and no pure dephasing, at any photon number, since the jump
terms of both Lindblad channels vanish in the coherent block. The two differ
only in the incoherent remainder, which loss removes from the waveguide and
dephasing returns to it.}. The drive is an $N$-photon Fock pulse,
$\ket{N_f}=(N!)^{-1/2}(A^\dag[f])^N\ket{0}$ with
$A^\dagger[f]=\int\!dt\,f(t)a^\dagger(t)$ and $f(t)$ a normalized mode, taken as a Gaussian
throughout with amplitude width $\sigma_t$,
$f(t)\propto e^{-t^2/2\sigma_t^2}$.

Pure dephasing can 
be modeled as classical multiplicative noise
$\xi(t)\sigma_{ee}$ with
$\braket{\xi}=0$ and $\braket{\xi(t)\xi(t')}=\gp\,\delta(t-t')$, equivalent on
averaging to the Lindblad term $\gp\mathcal{D}[\sigma^+\sigma^-]$. 
Per noise realization the dynamics are unitary and the state is pure, so every
object may be constructed for a pure state and averaged at the end; a
superscript $(r)$ denotes a quantity evaluated in one such realization.
Writing $\Ex[\,\cdot\,]$ for the
average over realizations, the distinction that organizes everything is whether
an observable is a first moment $\Ex[A]$ or a second moment $\Ex[A^*A]$, since
in general these are not related by $\Ex[A^*A]=|\Ex[A]|^2$.
 
For a Fock state $\braket{a_{\rm in}}=0$, so
no ratio
$\braket{a_{\rm out}}/\braket{a_{\rm in}}$ exists and the phase must be defined
interferometrically. The measurable object is the 
{\it cross-arm correlator}
$\Gba(t_1,t_2)=\braket{b^\dagger(t_1)b^\dagger(t_2)a(t_2)a(t_1)}$, with $b$ a field operator for the
reference arm that bypasses the emitter [Fig.~\ref{fig:schematic}(a)]. Since
that arm carries no dephasing noise, the average factorizes,
$\Gba(t_1,t_2)\propto f^*(t_1)f^*(t_2)\,\Ex[f'^{(r)}(t_1,t_2)]$, a first moment: no factorization is
needed, and the result is exact under dephasing with the single substitution
$\zeta\to\zeta(\gp)$.

A Fock pulse enters a waveguide containing a single emitter subject to pure
dephasing at rate $\gp$ [Fig.~\ref{fig:schematic}(a)]; the coupling is chiral,
so the emitter radiates only forward and there is no reflected field to account
for. Panels (b) and (c) contrast one and two photons under otherwise identical
conditions. In both cases the output separates into a coherent part, drawn
inverted because it carries the $\pi$ shift, and an incoherent remainder (dashed curve).
Increasing $\gp$ transfers weight from the first to the second while leaving the
sign of the coherent part untouched; this is the central result, made
quantitative below. The two rows differ in how fast that transfer happens, which
is the second result: the coherent two-photon amplitude survives to larger $\gp$
than the one-photon amplitude does. Photon number is conserved throughout; what
is lost is phase coherence with the drive, not photons.

\textit{Single-photon response.}---For the chiral emitter system, the coherent single-photon transmission is 
\begin{equation}
t(\omega)=1-\frac{\gamma}{\Gamma_2/2-i(\omega-\omega_0)}
=\frac{(\gp-\gamma)/2-i(\omega-\omega_0)}{(\gp+\gamma)/2-i(\omega-\omega_0)},
\label{eq:tw}
\end{equation}
reproducing the result of Ref.~\cite{Matias_2026} and, at $\gp=0$, the ideal
chiral result $|t|=1$, $\theta=\pi+2\arctan(2(\omega-\omega_0)/\gamma)$. On resonance
$t(\omega_0)=(\gp-\gamma)/(\gp+\gamma)$ is \emph{real}, so $\theta(\omega_0)$
can only take the values $0$ or $\pi$: it is a sign, not an angle, and a sign
cannot rotate. It is $\pi$ for all $\gp<\gamma$, undefined at $\gp=\gamma$ where
$t$ vanishes, and $0$ above.
We stress that $t(\omega)$ is the \emph{coherent} amplitude, a
first moment, and its vanishing is {\it not a loss of photons}: the geometry is chiral
so nothing is reflected, pure dephasing scatters nothing out of the waveguide,
 and the transmitted
flux obeys $\int n_{\rm out}dt=N$ exactly at every $\gp$,
a statement about the
signal arm alone, with no interferometer present.
At $\gp=\gamma$ the
input and the antiphase re-radiated field cancel exactly, so the photons still
emerge but with their phase scrambled, in the incoherent 
component; a photon
counter registers them, an interferometer does not.

Physically, one has $t=1+s$ with $s(\omega_0)=-2\gamma/(\gamma+\gp)$ the
coherently re-radiated field. 
Dephasing is phase diffusion: it damps the emitter coherence isotropically,
adding to the decay rate $\mathrm{Re}\,\zeta=(\Gamma+\gp)/2$ without
contributing anything imaginary, so it shrinks $|s|$ without rotating it. The $\pi$ is an \emph{overshoot}; at $\gp=0$ the
emitter radiates $|s|=2$ against an input of $1$ and the sum flips sign. While
the antiphase amplitude still exceeds the input the sum stays flipped; at
$|s|=1$ the two cancel exactly and there is no coherent field left to carry a
phase.

\begin{figure*}[t]
\centering
\includegraphics[width=\textwidth]{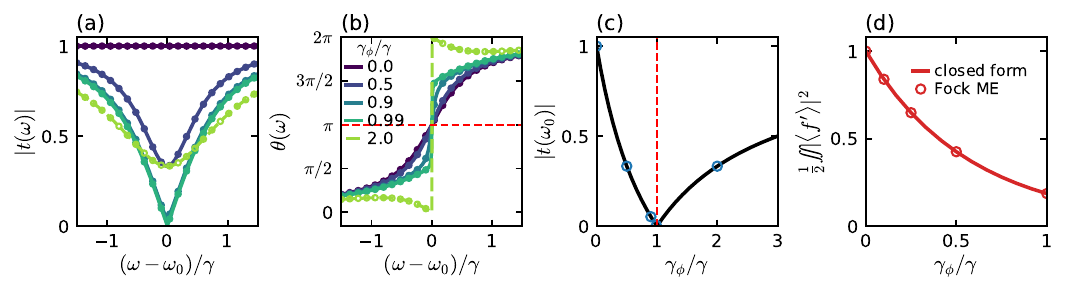}
\caption{Collapse of the coherent 
amplitude: single-photon (a)--(c), two-photon
(d). Throughout, lines are the closed-form results and symbols the independent
Fock-state master equation (Fock ME) of the End Matter. (a) Transmission
amplitude $|t(\omega)|$ and (b) phase $\theta(\omega)$ from Eq.~\eqref{eq:tw},
for $\gp/\gamma=0,\,0.5,\,0.9,\,0.99,\,2$, with circles the $N=1$ Fock ME
(agreement better than $4\times10^{-4}$); the dashed curve is the single above-threshold
case $\gp=2\gamma$, where $\theta(\omega_0)$ has flipped to $0$.
$\theta(\omega_0)=\pi$ exactly for every $\gp<\gamma$. The slope at resonance, $d\theta/d\omega|_{\omega_0}=2/(\gamma+\gp)+2/(\gamma-\gp)$,
diverges at $\gp=\gamma$ where $|t(\omega_0)|$ 
vanishes---the $\pi$ is then
carried by an amplitude of zero. (c) $|t(\omega_0)|$ against $\gp$. (d) Coherent
fraction of the two-photon amplitude, 
$\tfrac12\iint|\Ex[f']|^2$:
the closed
form Eq.~\eqref{eq:fp} against the $N=2$ Fock ME, which agree to
$4\times10^{-4}$ across the range.}
\label{fig:phase}
\end{figure*}

The distinction is between first and second moments, since Eq.~\eqref{eq:tw} is a
first moment and can cancel, whereas photon number is a second moment and cannot.
The re-emitted light appears as the incoherent Lorentzian of
Ref.~\cite{Matias_2026}. Figure~\ref{fig:phase} shows the resulting collapse.

\textit{Closed-form two-photon amplitude.}---Since $\Gba$ is linear in the
scattered amplitude, the two-photon phase needs only single amplitudes of the
form $\bra{0}\cdots\ket{\psi}$, with the vacuum already in the bra. For these the
intervals over which the emitter accumulates a random phase are disjoint, so the
noise average can be taken exactly, with no factorization.

The two-photon output amplitude is
$f'(t_1,t_2)=\bra{0}a_{\rm out}(t_2)a_{\rm out}(t_1)\ket{2_f}$, the probability
amplitude for detecting one transmitted photon at $t_1$ and a second at $t_2$.
Substituting the input--output relation for each $a_{\rm out}$ generates four
contributions, labelled by how many of the two photons have interacted with the
emitter,
\begin{equation}
f'=X_0-i\sqrt\gamma\,(X_1+X_2)-\gamma X_3 ,
\label{eq:Xsplit}
\end{equation}
where $X_0=\sqrt2 f(t_1)f(t_2)$ is the amplitude for both photons to pass
through untouched, $X_1$ and $X_2$ carry one emitter operator each (one
photon scattered, one not) and
$X_3=\bra{0}\sigma^-(t_2)\sigma^-(t_1)\ket{2_f}$ is the term in which both are
scattered. Solving the hierarchy for each and collecting terms gives
\begin{equation}
f'(t_1,t_2)=\sqrt2\Big[\underbrace{\psi'(t_1)\psi'(t_2)}_{\rm linear}
+\underbrace{\gamma\,e^{-\zeta|t_2-t_1|}C^{(1)}(t_{\min})^2}_{\rm nonlinear}\Big],
\label{eq:fp}
\end{equation}
where $t_{\min}=\min(t_1,t_2)$, \mbox{$\psi'(t)=f(t)-i\sqrt\gamma\,C^{(1)}(t)$} is the
transmitted single-photon amplitude, and
\mbox{$C^{(1)}(t)=-i\sqrt\gamma\int_{-\infty}^{t}\!ds\,e^{-\zeta(t-s)}f(s)$} is the emitter
coherence, which carries the dephasing through $\zeta$.
Equation~\eqref{eq:fp} is the central analytic result: an $S$-matrix-like
transfer amplitude for a dephasing node, which can be chained to describe several
nodes in sequence, valid for arbitrary $\gp$ and arbitrary pulse shape,
including entangled and spectrally shaped inputs~\footnote{Equation~\eqref{eq:fp} is written for both photons in one mode. For
the input $A^\dagger[f_1]A^\dagger[f_2]|0\rangle$ with distinct envelopes $f_1,f_2$, the same
derivation gives
$f'(t_1,t_2)=\psi'_1(t_1)\psi'_2(t_2)+\psi'_2(t_1)\psi'_1(t_2)
+2\gamma e^{-\zeta|t_2-t_1|}C^{(1)}_1(t_{\min})C^{(1)}_2(t_{\min})$, with
$\psi'_j=f_j-i\sqrt\gamma C^{(1)}_j$ and $C^{(1)}_j$ driven by $f_j$; Eq.~\eqref{eq:fp}
follows on setting $f_1=f_2=f$ and dividing by $\sqrt2$, the normalization of
$|2_f\rangle$. Since the amplitude is linear in the input, entangled two-photon inputs
follow by superposition. Nothing in the dephasing average depends on the input,
so the $\gp$ dependence is unchanged.}.

Validation is threefold, and detailed in the End Matter. First, against a Fock-state
master equation solved with no factorization, the relative error is
$4.3\times10^{-5}$ at $\gp=0$ rising only to $3.8\times10^{-4}$ at $\gp=2\gamma$.
Second, against frequency-domain scattering theory at $\gp=0$, the worst disagreement
over thirteen points is $1.7\times10^{-4}$. 
Third, at $\omega_1=\omega_2=\omega_0$ the closed form collapses to a single
quadrature that needs no time or frequency grid, giving $f'/f=-1.462761$ against
$-1.4628$ from scattering theory.
The nonlinear term is not a correction: it reaches
$2.0$--$3.4$ times the linear one, which is why the two-photon response is nothing
like $t(\omega_1)t(\omega_2)$.

\begin{figure*}[t]
\centering
\includegraphics[width=\textwidth]{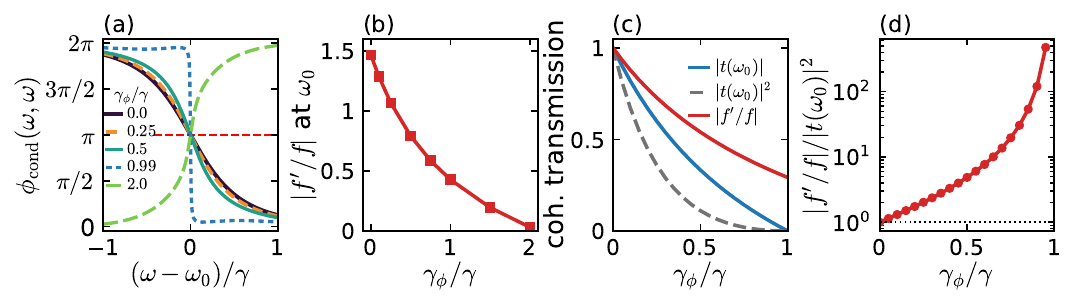}
\caption{The conditional phase across frequency and dephasing, and the unequal attenuation of the two channels.
(a) Conditional phase along the two-photon diagonal, $\phi_{\rm cond}(\omega,\omega)$, unwrapped through $\pm\pi$, for
$\gp/\gamma=0$ (solid), $0.25$ (orange dashed), $0.5$, $0.99$ (dotted) and $2$ (dashed, above threshold).
At $\omega_0$ it is exactly $0$ or $\pi$ for any dephasing rate~\cite{note:quantized_phase} and equals $\pi$ at every rate examined; it is
antisymmetric about $\pi$, so a symmetric resonant pulse returns $\pi$ on
averaging. Its width is set by the linear reference $t(\omega)^2$, which vanishes at $\omega_0$ as $\gp\to\gamma$; the sweep
narrows to $|\gamma-\gp|/2$, is undefined at $\gp=\gamma$, and reverses above threshold. (b) Resonant two-photon transmission
$|f'/f|$ at $\omega_1=\omega_2=\omega_0$ (squares, Fock-state master equation). (c) Resonant coherent transmission of the
one-photon channel, $|t(\omega_0)|$, and the two-photon channel, $|f'/f|$, each normalized to its $\gp=0$ value, against the
independent-scattering expectation $|t(\omega_0)|^2$. (d) Ratio of the two-photon transmission to that expectation, normalized
as in (c), the factor by which the two-photon channel outlives two independent single-photon scatterings.
}
\label{fig:cond}
\end{figure*}

\textit{Conditional phase.}---A $c$-phase gate is characterized by the phase of the two-photon amplitude
relative to 
independent single-photon scattering,
\begin{align}
R_{\rm cond}(t_1,t_2)&=\frac{f'(t_1,t_2)}{\sqrt2\,\psi'(t_1)\psi'(t_2)}\nonumber\\
&=1+\frac{\gamma\, e^{-\zeta|t_2-t_1|}C^{(1)}(t_{\min})^2}{\psi'(t_1)\psi'(t_2)},
\label{eq:cond}
\end{align}
with $\phi_{\rm cond}=\arg R_{\rm cond}$. The entire conditional phase comes
from the bound-state term. On resonance with real $f$ everything in
Eq.~\eqref{eq:cond} is real, so $\phi_{\rm cond}\in\{0,\pi\}$ exactly~\cite{note:quantized_phase}, and we find $\phi_{\rm cond}=\pi$ at every dephasing rate examined,
$\gp/\gamma=0$ to $2$, for our chosen input pulse [Fig.~\ref{fig:cond}(a)]~\footnote{While quantization of $\phi_{\rm cond}$ is guaranteed by resonance and real $f$, whether $R_{\rm cond}$ is positive or negative is dependent on the strength of the nonlinear interaction.}. The point $\gp=\gamma$ is
excluded there only because the linear reference $t(\omega_0)$ vanishes, making
the ratio singular even though $f'$ is finite. The value $\pi$ requires the
bound-state term to exceed $\psi'(t_1)\psi'(t_2)$. For a Gaussian pulse at a pulse width of $\gamma\sigma_t=1$ this
holds out to $\gp\simeq2\gamma$ and fails only once the emitter has decoupled,
where the node is in any case nearly transparent.

The penalty appears entirely in the amplitude. The resonant two-photon
transmission $|f'/f|$ falls monotonically, $1.46\to1.07\to0.79\to0.43\to0.03$
for $\gp/\gamma=0,\,0.25,\,0.5,\,1,\,2$ [Fig.~\ref{fig:cond}(b)], and the
coherent fraction 
$\tfrac12\iint|\Ex[f']|^2$
falls from $1$ to $0.19$ over
$\gp=0\to\gamma$~[\cref{fig:phase}(d)] while the photon number stays exactly $2$. For a $c$-phase gate
the figure of merit is $\phi_{\rm cond}=\pi$ with as much transmission as
possible; the phase condition is met exactly across the relevant range, and the
whole cost is contrast.

That the {\it conditional phase survives} is consistent with the single-photon
inversion above. The linear phase enters the two-photon observable twice, so the
flip $\theta(\omega_0):\pi\to0$ at $\gp=\gamma$ changes $\theta+\theta$ by $2\pi$, that is, not at all---and $R_{\rm cond}$, which divides out
$t(\omega_1)t(\omega_2)$, is invariant under a sign flip common to both photons.
A $c$-phase gate is a difference between one- and two-photon responses, so
whatever is common to both cancels from it. That insensitivity is exactly what
makes $\phi_{\rm cond}$ robust, and exactly why the penalty has nowhere to go
except into the amplitude.
Continuous-wave chiral nonlinear optics and all-optical control draw on the same
interaction with coherent fields rather than Fock-state
pulses~\cite{Dufresne2026a,Dufresne2026b}.

\textit{Dephasing distorts rather than scales.}---It is tempting to summarize the above as an overall scale, writing the realized
node map as $V=\eta\,U$ with $U$ the target unitary and $\eta\le1$ real. A
normalized, phase-sensitive fidelity would then divide out $\eta$ and report
$F=1$ identically, and the dephasing penalty would be pure loss: a rate
multiplying the heralding probability, compounding as $\eta^N$ over $N$ cascaded
nodes.

Importantly, no single $\eta$ exists, and Fig.~\ref{fig:cond}(c,d) shows why.
The one- and
two-photon channels are suppressed at different rates. Normalized to their
$\gp=0$ values, the single-photon transmission falls as
$|t(\omega_0)|=(\gamma-\gp)/(\gamma+\gp)$, reaching zero at $\gp=\gamma$, while
the resonant two-photon transmission falls only to $0.29$ of its $\gp=0$ value
there (i.e.,\ $|f'/f|=0.43$). Were the two
photons scattering independently the latter would track $|t(\omega_0)|^2$;
instead it exceeds that expectation by factors of $1.7$, $3.3$ and $7.7$ at
$\gp/\gamma=0.2,\,0.4,\,0.6$, and diverges from it at $\gp=\gamma$, where the
node transmits two-photon amplitude while transmitting none at one photon.
A single scalar $\eta$ cannot describe this. 

Dephasing does two distinct things:
a part common to all sectors, which is a pure scale, and a part that reweights
the coherent photon-number sectors against one another, which is a genuine distortion of
the target unitary. Only the first is invisible to a normalized fidelity.
{\it Neither is trainable}: a network tunes phase shifters and splitting ratios, i.e.,
unitary parameters, whereas the missing weight has left the coherent subspace
irreversibly into incoherent emission, and the reweighting is fixed by
$\gp/\gamma$ at the emitter, a rate over which the circuit has no handle. For
architectures that cascade emitter
nonlinearities~\cite{Ewaniuk2023,Ewaniuk2025,QPNN2026}, 
this is precisely the imperfection the central mechanism cannot address, and
it should be accounted for as two independent penalties, not a single efficiency factor. A network-level study will be presented
in another work.

The reweighting is directly measurable, and with apparatus that exists. Both
$|t(\omega_0)|$ and $|f'/f|$ are coherent amplitudes referenced against an
unscattered arm, i.e., precisely the quantity the Mach--Zehnder measurement of
Ref.~\cite{Staunstrup2024} measures. Sending first a one-photon and then
a two-photon pulse through the same interferometer, on resonance and at fixed
$\gp$, and comparing the two-photon fringe contrast against the square of the
one-photon contrast tests Fig.~\ref{fig:cond}(d) directly: a linear element, or
an emitter acting on the two photons independently, gives unity, whereas the
bound-state term gives the growing excess predicted here. No absolute
calibration is needed, since both are normalized to their own $\gp=0$ values.


\textit{The standard factorization fails.}---The two-photon
\emph{intensity} correlator, \mbox{$\Gtt=\braket{a^\dagger_{\rm out}(t_1)a^\dagger_{\rm
out}(t_2)a_{\rm out}(t_2)a_{\rm out}(t_1)}$}, is a different object from the phase
observable above, and here the situation is worse. It is a second moment: both
operators carry the dephasing noise, so there is nothing noise-free to factorize
against. The standard construction inserts a vacuum projector between the
emitter operators, which replaces an average of a product by a product of
averages. That step is exact when the zero-excitation subspace is one
dimensional; this is true for arbitrary optical loss, which is why the published
$\gp=0$ scattering theory is unaffected, but a Markovian $\gp$ requires bath
weight at zero frequency, and the dephasing sector is never vacuum.

The factorization fails outright, since it breaks a conservation law.
Pure
dephasing destroys no excitations, so $\iint\Gtt\,dt_1dt_2=N(N-1)$ exactly at
every $\gp$; the factorized expression instead falls short by $16$--$81\%$ over
$\gp/\gamma=0.1$--$1$, the window in which the theory is wanted
[Fig.~\ref{fig:gate}(c), End Matter]. The discarded weight is a covariance,
$\Ex[|f'|^2]-|\Ex[f']|^2$, hence one-signed: the factorization always
undercounts, and the error is first order in $\gp$, so there is no safe
weak-dephasing regime to retreat to. The missing pairs are not missing from the
experiment---they are the incoherently re-emitted photons of
Fig.~\ref{fig:schematic}(b,c), which a photon counter records and an interferometer
does not. Per noise realization the output state is pure, so the exact
replacement is a single line, $\Gtt=\Ex[|f'^{(r)}|^2]$ with $f'^{(r)}$ the
per-realization amplitude; the End Matter gives it in closed form and validates
it pointwise.

\textit{Conclusions.}---We have derived the two-photon scattered amplitude of a chirally coupled two-level
emitter with pure dephasing in closed form, Eq.~\eqref{eq:fp}, 
validated against
an independent Fock-state master equation 
and against
frequency-domain scattering theory at $\gp=0$. 
The conditional phase is pinned at $\pi$ for every pure-dephasing rate examined, so \emph{pure dephasing destroys the two-photon
amplitude rather than rotating the nonlinear phase}.
The one- and two-photon channels are attenuated unequally, so
the dephased node does not simply pass less light: it changes the balance
between the one- and two-photon channels, an error that a normalized fidelity
does not register and that no phase setting can repair.

Two directions follow. First, chaining Eq.~\eqref{eq:fp} over several nodes would give network-scale predictions without
re-simulating the many-body dynamics, provided the cascade closes analytically even though each node sees an input already
correlated by the previous one. Second, the white-noise model of pure dephasing, shared with all the cited work, is justified here.
Phonon dephasing in quantum dots is non-Markovian~\cite{PhysRevLett.104.017402,PhysRevLett.106.247403,IlesSmith2017,PhysRevLett.118.253602,PhysRevB.98.045309,Denning2020},
but its correlation time, a few picoseconds, is far below the pulse durations of interest. At $\gamma\sigma_t\sim1$ the acoustic
sideband then reduces to a temperature-dependent Franck--Condon factor $\langle B\rangle^2$ ($\langle B\rangle^4$ at two photons), a
calculable prefactor that spectral filtering largely removes~\cite{Dusanowski2022,Denning2020}. Non-Markovian corrections matter
mainly under strong driving, where pulse shaping can suppress them~\cite{PhysRevB.98.045309,PhysRevLett.128.167403,Crowder2026}.



\begin{acknowledgments}
We acknowledge support from the Natural Sciences and Engineering Research
Council of Canada (NSERC), the Canadian Foundation for Innovation
(CFI), and Queen's University, Canada.
\end{acknowledgments}

\bibliographystyle{apsrev4-2}
\bibliography{refs_merged,notes}

\clearpage
\newpage

\clearpage
\appendix
\section*{End Matter}

\textit{Fock-state master equation (Fock ME) benchmark.}---No scattering treatment includes
dephasing, so there is nothing in the literature to check Eq.~\eqref{eq:fp}
against. We therefore validate it against a Fock-state master equation solved
under identical assumptions. For an $N$-photon Fock pulse with coupling operator
$L=-i\sqrt{\gamma_R}\,\sigma^-$, define generalized operators $\rho_{m,n}$,
$m,n=0\ldots N$, obeying
\begin{align}
\dot\rho_{m,n}&=\mathcal{L}\rho_{m,n}+\sqrt m\,f(t)\,[\rho_{m-1,n},L^\dagger]
\nonumber\\
&\quad+\sqrt n\,f^*(t)\,[L,\rho_{m,n-1}],\label{eq:fockme}\\
\mathcal{L}\rho&=-i[\Delta\sigma_{ee},\rho]+\Gamma\mathcal{D}[\sigma^-]\rho
+\gp\,\mathcal{D}[\sigma^+\sigma^-]\rho,
\end{align}
with $\mathcal{D}[O]\rho=O\rho O^\dagger-\tfrac12\{O^\dagger O,\rho\}$ and
$\rho_{m,n}(t_0)=\ket{g}\bra{g}$; the physical state is
$\rho_{N,N}$~\cite{Gheri1998,BaragiolaCombes2017,Combes2017TheNetworks,Kiilerich2019}.
Dephasing enters as one more Lindblad term: no factorization, no vacuum
insertion, no restriction on $\gp$. The labels $(m,n)$ count the photons
dressing ket and bra, so the amplitudes entering the phase live in off-diagonal
blocks; $f'=\bra{0}\cdots\ket{2_f}$ has a two-photon ket and a vacuum bra, hence
$\braket{f'(t_1,t_2)}=\mathrm{Tr}\{[\mathcal{A}\,\mathcal{E}_{t_1\to
t_2}[\mathcal{A}\rho]]_{2,0}\}$, with
$(\mathcal{A}\rho)_{m,n}=L\rho_{m,n}+\sqrt m f(t)\rho_{m-1,n}$ and
$\mathcal{E}_{t_1\to t_2}$ propagation by Eq.~\eqref{eq:fockme}.

At $\gp=0$ the Fock ME reproduces the benchmarks of Ref.~\cite{p694-46xh},
including the peak $|G^{(1)}_{N}|=0.2075$ and $0.5207$ for $N=1,2$, the node at $\gamma t=4.00$, emitter population peaks $0.571$ and $0.798$, the flux
identity $n_L=\gamma n_{\rm TLS}/2$, and the bimodal-pulse results of
Ref.~\cite{sofia2025}. Sum rules $\int n_{\rm out}=N$ and $\iint
G^{(2)}=N(N-1)$ hold to $10^{-5}$ at every $\gp$. Table~\ref{tab:valid} collects
the comparisons; each entry is the largest relative deviation between the closed
form and the reference calculation.

\begin{table}[b]
\caption{Validation of the closed-form two-photon amplitude.}
\label{tab:valid}
\begin{ruledtabular}
\footnotesize
\begin{tabular}{ll}
route (regime) & agreement \\ \colrule
Fock ME (time domain, all $\gp$) & $4.3\times10^{-5}$--$3.8\times10^{-4}$\\
analytic integral ($\omega_0$, grid free) & 5 sig.\ figures\\
scattering theory ($\gp=0$) & worst $1.7\times10^{-4}$\\
covariance core $M$ (Monte Carlo) & $3\times10^{-3}$ (MC floor)\\
2D $\Gtt$ map (cells $>2\%$ of peak) & $4.3\times10^{-4}$ ($\gp=0$)
\end{tabular}
\end{ruledtabular}
\end{table}

\textit{Derivation of the two-photon amplitude.}---Projecting the
Heisenberg--Langevin equation for $\sigma^-$ onto Fock sectors closes a
hierarchy labelled by the photon numbers of bra and ket. For one photon,
\begin{equation}
\dot C^{(1)}=-\zeta\,C^{(1)}-i\sqrt\gamma\,f(t)\big[2P^{(0)}(t)-1\big],
\label{eq:C1}
\end{equation}
with $C^{(1)}=\bra{0}\sigma^-\ket{1_f}$ and $P^{(0)}$ the initial population;
for an emitter starting in $\ket{g}$ this integrates to the $C^{(1)}(t)$ quoted
below Eq.~\eqref{eq:fp}. All the dephasing dependence sits in $\zeta$: each
$\sigma^-(t)$ carries $e^{-i\varphi(t)}$ with
$\varphi(t)=\int_{t_0}^{t}\!ds\,\xi(s)$, only the difference
$\varphi(t)-\varphi(s)$ over the interval separating excitation from
de-excitation appears, and the Gaussian average gives $e^{-\gp(t-s)/2}$.

Each $X_i$ in Eq.~\eqref{eq:Xsplit} is a single amplitude with the vacuum
already in the bra, so no factorization is performed and the average is exact.
The hierarchy closes from the top: $X_0$ is known outright, $X_1,X_2$ are driven
by it, and $X_3$ by those. Substituting back and separating terms that factorize
as a function of $t_1$ times a function of $t_2$ from the one term that does not
yields Eq.~\eqref{eq:fp}: the factorizable part is $\psi'(t_1)\psi'(t_2)$, two
independent scatterings, and the remainder is the bound-state term, whose
exponential is the emitter coherence surviving between the two events.

\textit{Why the standard factorization fails.}---Expanding
$\Gtt=\braket{a^\dagger_{\rm out}(t_1)a^\dagger_{\rm out}(t_2)a_{\rm
out}(t_2)a_{\rm out}(t_1)}$ gives $2^4=16$ terms, according to whether each
operator is input field or emitter; the emitter-bilinear ones are conventionally
factorized by inserting a vacuum projector between them---the ``vacuum
cut'', so called because between the two emitter operators the system must
return to the zero-excitation subspace. That is exact whenever that subspace is
one dimensional.
Loss reservoirs are vacuum to excellent approximation, so the cut survives arbitrary
$\gamma_0$; we verify a population defect of $6\times10^{-10}$ at
$\gamma_0=2\gamma_R$. A Markovian $\gp\neq0$, however, needs bath weight at zero
frequency, and the dephasing sector is \emph{never} vacuum.

\begin{figure*}[t]
\centering
\includegraphics[width=\textwidth]{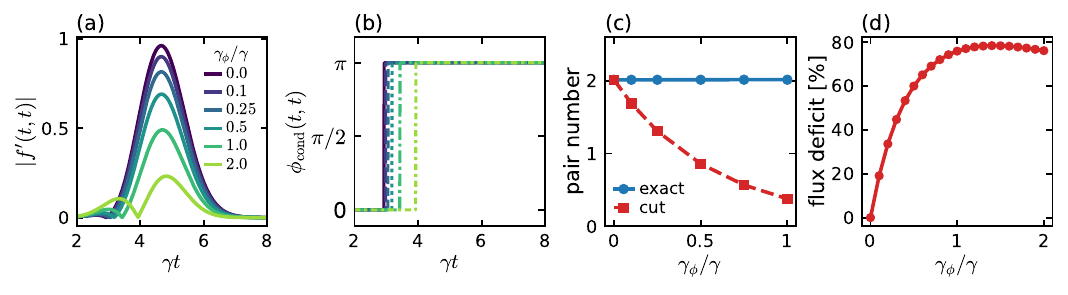}
\caption{Panels (a)--(c) are $N=2$, panel (d) is $N=1$. (a) Equal-time two-photon amplitude $|f'(t,t)|$: lines the closed form,
Eq.~\eqref{eq:fp}, circles the Fock ME.
(b) The conditional phase sits at $\pi$ wherever the amplitude is appreciable,
at every $\gp$; the curves are offset only by the crossing time,
$\gamma t=2.93,\,2.98,\,3.05,\,3.18,\,3.43,\,3.93$ in order of increasing $\gp$.
The step on the leading edge is a sign change of a real quantity passing through
zero, not a rotation. (c) 
Integrated pair number: the exact value $N(N-1)=2$, reproduced as $2.013$ at
every $\gp$, against the vacuum cut $\iint|\Ex[f']|^2$, which falls short by up
to $81\%$. 
(d) The single-photon counterpart: the fraction of the transmitted
photon flux the cut fails to account for. It is non-monotonic because strong
dephasing decouples the emitter: $C^{(1)}\to0$, so the transmitted field
approaches the unscattered input and the factorization becomes exact again. The
deficit peaks near $\gp\simeq1.4\gamma$ and decays; the pair deficit of (c)
behaves the same way, with its maximum near $\gp\simeq2\gamma$, beyond the range
plotted.}
\label{fig:gate}
\end{figure*}

The correct object is a \emph{covariance}. Take the all-emitter term, the only one
needing the insertion on both sides. Within a realization the state is pure and
the insertion exact, so it is
$\Ex[X_3^{*(r)}(t_1,t_2)X_3^{(r)}(t_1,t_2)]$, whereas inserting the projector
after averaging gives $\Ex[X_3^*]\Ex[X_3]$. The difference is the covariance of
$X_3$ across realizations, nonzero because both factors are driven by the same
$\xi(t)$. The Gaussian average can be done exactly, giving a quadruple integral
dressed by $e^{\gp\Omega}$, with $\Omega$ the total time over which a raising and
a lowering leg co-propagate; $\Omega\to0$ recovers the cut. Two consequences
follow that numerics alone would not give: the error is first order,
$e^{\gp\Omega}-1\simeq\gp\Omega$, so there is no safe weak-dephasing regime; and
the cut is a strict lower bound: $\Omega\ge0$ off the diagonal, and on it
the exact term is $\Ex[|X_3|^2]$, which cannot fall below $|\Ex[X_3]|^2$ by Cauchy--Schwarz.

The sixteen terms are not in fact needed separately. Per realization the output
state is pure, so the exact correlator and its factorized counterpart are
\begin{equation}
\Gtt(t_1,t_2)=\Ex\!\left[|f'^{(r)}(t_1,t_2)|^2\right],\ \
\Gttfac=\left|\Ex[f'^{(r)}]\right|^2,
\label{eq:allT}
\end{equation}
and Eq.~\eqref{eq:allT} supersedes the term-by-term construction: the sixteen
terms are just $|f'^{(r)}|^2$ sorted by powers of $a_{\rm in}$ versus $\sigma^-$.
The simplest case is $N=1$, where the difference between them reduces to the
population defect $\Delta_P=P^{(1)}-|C^{(1)}|^2=
\gp\int_{-\infty}^{t}\!ds\,e^{-\Gamma(t-s)}|C^{(1)}(s)|^2\ge0$, a variance,
already published as $Q(t)$ in Ref.~\cite{Matias_2026}, App.~C. We claim no
novelty at $N=1$; the point is that the variance reading makes the sign manifest
and generalizes.

\textit{Pulse duration.}---The duration used throughout, $\gamma\sigma_t=1$, is
close to the best available choice, and two quantities set it.
Gate performance is known to depend strongly on
pulse bandwidth~\cite{Nysteen2017,Nysteen2015}; dephasing supplies a
second, competing constraint. The bound-state share of $\tfrac12\iint|f'|^2$
peaks at $\gamma\sigma_t\simeq1.5$--$2$: a shorter pulse is broader than the
linewidth, so most spectral weight sits where $t(\omega)\to1$, while a longer one
weakens the nonlinearity relative to the linear term. 
Dephasing robustness, by contrast, favors short pulses monotonically, since the
damage accumulates with the time the emitter holds coherence: the coherent fraction at
$\gp=0.5\gamma$ falls $0.63,\,0.49,\,0.43,\,0.38,\,0.27$ at
$\gamma\sigma_t=0.25,\,0.5,\,1,\,2,\,4$.
At $\gp=0$ it is unity at every
duration, as unitarity requires.

\textit{The noise model.}---White-noise pure dephasing, $\gp\mathcal{D}[\sigma^+\sigma^-]$, is an idealization shared here by the input/output theory, the Fock-state master equation, and the cited literature
alike, and its adequacy rests on a separation of timescales. The phonon correlation time is a few ps, short against the pulse
durations of interest, so at $\gamma\sigma_t\sim1$ the acoustic sideband does not
reshape the coherent response but multiplies it by a temperature-dependent
Franck--Condon factor $\langle B\rangle^2$, becoming $\langle B\rangle^4$ at two
photons; this is a known prefactor rather than a structural change, most of which
spectral filtering removes~\cite{Dusanowski2022,Denning2020}. What remains is
small at picosecond durations and matters chiefly under intense driving, where
pulse shaping suppresses it
further~\cite{PhysRevB.98.045309,PhysRevLett.128.167403,Crowder2026}. Pure
dephasing is thus the dominant channel here, which is why treating it exactly is
the useful step.

Slow noise is a different matter. Spectral diffusion (the wandering of the
emitter's transition frequency as nearby charges rearrange) has a correlation
time of microseconds to milliseconds, so each repetition sees an essentially
static detuning drawn from a distribution rather than a fluctuating one. The
measured signal is then an average over $\Delta$, not a larger $\gp$, which
matters for any interferometric measurement accumulated over many shots, where
spectral wandering broadens the line without being pure dephasing at all. It is
also straightforward to include, since Eq.~\eqref{eq:fp} holds at arbitrary
detuning: the observable follows by averaging over the detuning distribution,
and the variance that white noise hides becomes directly measurable because
different runs genuinely sample different detunings. Like pure dephasing, this
costs amplitude rather than adding a phase error, and no phase setting recovers
it.

\end{document}